\documentclass[journal,twocolumn,10pt]{IEEEtran}
\IEEEoverridecommandlockouts 
\usepackage{epsfig,latexsym}
\usepackage{float}
\usepackage{indentfirst}
\usepackage{amsmath}
\usepackage{amssymb}
\usepackage{times}
\usepackage{subfigure}
\usepackage{pifont}
\usepackage{psfrag}
\usepackage{cite}
\usepackage{url}
\usepackage{lastpage}
\usepackage{epstopdf}
\usepackage{bm}
\usepackage{color}
\usepackage{fancyhdr}
\usepackage[center]{caption2}
\usepackage{balance}
\usepackage{fancyhdr,lastpage}
\usepackage{hyperref}

\begin{document}
\title{Robust Sum Secrecy Rate Optimization for MIMO Two-way Full Duplex Systems}
\vspace{1em}
\author{\IEEEauthorblockN{Zheng Chu$^{1}$, Tuan Anh Le$^{1}$, Huan X. Nguyen$^{1}$, Arumugam Nallanathan$^{2}$, and Mehmet Karamanoglu$^{1}$} \\
	\vspace{0.5em}
$^{1}$	Faculty of Science and Technology, Middlesex University, London, United Kingdom. \\
$^{2}$ Department of Informatics, King’s College London, London, United Kingdom.\\
\vspace{0.5em}
(Email: z.chu@mdx.ac.uk)
}
\vspace{-0.2in}
\maketitle
\thispagestyle{empty}
\begin{abstract}
This paper considers multiple-input multiple-output (MIMO) full-duplex (FD) two-way secrecy systems. Specifically, both multi-antenna FD legitimate nodes exchange their own confidential message in the presence of an eavesdropper. Taking into account the imperfect channel state information (CSI) of the eavesdropper, we formulate a robust sum secrecy rate maximization (RSSRM) problem subject to the outage probability constraint of the achievable sum secrecy rate and the transmit power constraint. Unlike other existing channel uncertainty models, e.g., \emph{norm-bounded} and \emph{Gaussian-distribution}, we exploit a moment-based random distributed CSI channel uncertainty model to recast our formulate RSSRM problem into the convex optimization frameworks based on a \emph{Markov's inequality} and robust conic reformulation, i.e., semidefinite programming (SDP). In addition, difference-of-concave (DC) approximation is employed to iteratively tackle the transmit covariance matrices of these legitimate nodes. Simulation results are provided to validate our proposed FD approaches.
\end{abstract}
\IEEEpeerreviewmaketitle
\setlength{\baselineskip}{1\baselineskip}
\newtheorem{definition}{Definition}
\newtheorem{fact}{Fact}
\newtheorem{assumption}{Assumption}
\newtheorem{theorem}{Theorem}
\newtheorem{lemma}{Lemma}
\newtheorem{corollary}{Corollary}
\newtheorem{proposition}{Proposition}
\newtheorem{example}{Example}
\newtheorem{remark}{Remark}
\newtheorem{algorithm}{Algorithm}
\section{Introduction}\label{section:Introduction}
With the recent development of self-interference cancellation (SIC), full-duplex (FD) communication has been considered as one of promising physical-layer techniques to satisfy the exponential growth in high data rate for the fifth generation (5G) mobile networks \cite{Kim_IEEE_COM_TUT_2015}. Particularly, unlike the traditional half-duplex (HD) transmission, FD doubly improves the spectral efficiency through transmission and reception simultaneously on the same frequency band.
Self-interference (SI) has been considered a major challenge that caused by the signal leakage from transmission to reception for FD nodes. SI can be partially cancelled through analogue circuits and digital signal processing, however, the residual SI still impairs the performance of FD systems if it is not properly controlled \cite{Bharadia_2013}. Recently, multi-antenna FD system has been developed to further enhance the spectral efficiency, specifically, \cite{Suraweera_TWC_FD_MIMO_2013} investigated the end-to-end outage probability of MIMO FD single-user relaying systems. In \cite{Derrick_TCOM_FD_MIMO_OFDMA_2012}, a resource allocation algorithm was proposed for the maximization of the end-to-end system data rate of multi-carrier MIMO FD relaying systems. Moreover, there is an increasing interests in various applications of FD communication, such as physical-layer security (PLS) \cite{Gan_Zheng_TSP_FD_Jam_2013,Feifei_Gao_TSP_Sec_FD_2014,Derrick_TWC_Sec_FD_2016,Renhai_Feng_TVT_Sec_FD_2016}.

PLS was developed based on an information-theoretic approach to provide information security at the physical-layer by exploiting the difference between the mutual information of the legitimate use and the eavesdropper \cite{Ma_Sig_Process_J11}. Recently, secure FD system have been seen as a promising paradigm to double spectrum and satisfy reliable transmission simultaneously. In \cite{Gan_Zheng_TSP_FD_Jam_2013}, the authors exploited the FD characteristics of the legitimate user to receive desired information and transmit jamming signal to interfere with the eavesdropper. In \cite{Feifei_Gao_TSP_Sec_FD_2014} and \cite{Derrick_TWC_Sec_FD_2016}, the authors exploited secrecy designs in cellular networks, where it consists of an FD base station (BS) and multiple HD uplink/downlink mobile users. For both of these works, the semidefinite programming (SDP) relaxation-based approach was employed to maximize the achievable downlink secrecy rate \cite{Feifei_Gao_TSP_Sec_FD_2014} or to minimize the uplink/downlink transmit powers under the achievable secrecy rate constraints \cite{Derrick_TWC_Sec_FD_2016}. Also, in \cite{Renhai_Feng_TVT_Sec_FD_2016}, robust transmit solution has been developed for two-way secure FD system based on the norm-bounded channel uncertainty model. However, using a single antenna at the receive side may lead to a fact that the performance is limited by the residual self-interference introduced by the imperfection of the transmit front-end chain \cite{Bharadia_2013}. On the other hand, it is not always possible to estimate the channel error bound or distribution at the legitimate nodes, thus, a novel channel uncertainty model will be considered. Both gaps motivate this paper.

In this paper, we investigate a MIMO secrecy two-way FD system, specifically, both multi-antenna FD legitimate nodes exchange their own confidential information in the presence of a multi-antenna eavesdropper. This paper takes into account the imperfection in the estimation of eavesdropper's channel state information (CSI). This paper differs from the existing channel uncertainty models, i.e., bounded-sphere \cite{Cuma_TVT_MIMO_Sec_2014} and Gaussian random \cite{Zheng_WCL_2015}, by investigating a \emph{moment-based} channel uncertainty model \cite{Qiang_Li_Distributionally_ICASSP_2014}, i.e., the first and second-order statistics of the channel errors are available whereas the exact distribution is not known. This channel uncertainty model is motivated by the fact that it is easier to estimate the error statistics than the exact error bound or distribution. We formulate a robust sum secrecy rate maximization (RSSRM) problem, subject to the outage probability constraint of the achievable sum secrecy rate and the transmit power constraint. Due to the non-convexity of the proposed RSSRM problem, we introduce two robust designs based on a \emph{Markov's inequality} and SDP to recast it into convex optimization frameworks. In addition, difference-of-concave (DC) approximation is employed to iteratively update the transmit covariance matrices of both legitimate nodes.
\section{System Model}\label{section:System_model}
In this section, we consider a secure MIMO FD system consisting of two multi-antenna legitimate nodes, named $ \mathcal{U}_1 $ and $ \mathcal{U}_2 $, who exchange their confidential information, and a multi-antenna eavesdropper, named $ \mathcal{E} $, who overhears these transmissions. It is assumed that both $ \mathcal{U}_1 $ and $ \mathcal{U}_2 $ work in FD mode, i.e., $ \mathcal{U}_1 $ ($ \mathcal{U}_2 $) transmits information to and receive from $ \mathcal{U}_2 $ ($ \mathcal{U}_1 $) simultaneously. Both $ \mathcal{U}_1 $ and $ \mathcal{U}_2 $ are equipped with $ N_{T,1} $ and $ N_{T,2} $ transmit antennas and $ N_{R,1} $ and $ N_{R,2} $ receive antennas, respectively, whereas $ \mathcal{E} $ consists of $ N_{E} $ antennas. The channel coefficients from the transmit antennas of $ \mathcal{U}_1 $ to the receive antennas of $ \mathcal{U}_2 $, $ \mathcal{E} $ and $ \mathcal{U}_1 $ can be denoted as $ \mathbf{H}_{12} \in \mathbb{C}^{N_{T,1} \times N_{R,2}} $, $ \mathbf{H}_{e,1} \in \mathbb{C}^{N_{T,1} \times N_{E}} $, and $ \mathbf{H}_{11} \in \mathbb{C}^{N_{T,1} \times N_{R,1}} $, respectively. While, $ \mathbf{H}_{21} \in \mathcal{C}^{N_{T,2} \times N_{R,1}} $, $ \mathbf{H}_{e,2} \in \mathbb{C}^{N_{T,2} \times N_{E}} $, and $ \mathbf{H}_{22} \in \mathbb{C}^{N_{T,2} \times N_{R,2}} $ are defined as the channel coefficients from the transmit antennas of $ \mathcal{U}_2 $ to the receive antennas of $ \mathcal{U}_1 $, $ \mathcal{E} $ and $ \mathcal{U}_2 $, respectively. Thus, the received signal at $ \mathcal{U}_1 $ and $ \mathcal{U}_2 $ can be, respectively, given by
\begin{subequations}
\begin{align}
\mathbf{y}_{1} &~= \mathbf{H}_{21}^{H}\mathbf{x}_{2} + \mathbf{H}_{11}^{H}\mathbf{x}_{1} + \mathbf{n}_{1}, \label{eq:y_1} \\
\mathbf{y}_{2} &~= \mathbf{H}_{12}^{H}\mathbf{x}_{1} + \mathbf{H}_{22}^{H}\mathbf{x}_{2} + \mathbf{n}_{2},\label{eq:y_2}
\end{align}
\end{subequations}
where $ \mathbf{n}_1 \sim \mathcal{CN}(0, \sigma_1^2 \mathbf{I}_{N_{R,1} \times N_{R,1}}) $ and $ \mathbf{n}_2 \sim \mathcal{CN}(0, \sigma_2^2 \mathbf{I}_{N_{R,2} \times N_{R,2}}) $ are circularly symmetric Gaussian noises. $ \mathbf{x}_{1} \in \mathbb{C}^{N_{T,1}\times 1} $ and $ \mathbf{x}_{2} \in \mathbb{C}^{N_{T,2} \times 1} $ are desired signal from $ \mathcal{U}_1 $ and $ \mathcal{U}_2 $, respectively. The transmit covariance matrices of $ \mathcal{U}_1 $ and $ \mathcal{U}_2 $ can be defined as $ \mathbf{Q}_{1} = \mathbb{E} \{ \mathbf{x}_{1}\mathbf{x}_{1}^{H} \} $ and $ \mathbf{Q}_{2} = \mathbb{E} \{ \mathbf{x}_{2}\mathbf{x}_{2}^{H} \} $ with $ \mathbf{Q}_{1} \succeq \mathbf{0} $ and $ \mathbf{Q}_{2} \succeq \mathbf{0} $. The second terms of \eqref{eq:y_1} and \eqref{eq:y_2} are the self-interference (SI) induced by the FD operation of $ \mathcal{U}_1 $ and $ \mathcal{U}_2 $, respectively. Despite exploiting \emph{a priori} knowledge of $ \mathbf{x}_{1} $ and $ \mathbf{x}_{2}$, the SI at $ \mathcal{U}_1 $ and $ \mathcal{U}_2 $ can only be suppressed due to high SI power and hardware limitations. Thus, the achievable rate at $ \mathcal{U}_1 $ and $ \mathcal{U}_2 $ can be, respectively, written as
\begin{align}
R_{1} = \log \left| \mathbf{I} + \bigg(\sigma_1^2 \mathbf{I} + \xi_1 \mathbf{H}_{11}^{H}\mathbf{Q}_{1}\mathbf{H}_{11} \bigg)^{-1} \mathbf{H}_{21}^{H}\mathbf{Q}_{2}\mathbf{H}_{21} \right|, \nonumber\\
R_{2} = \log \left| \mathbf{I} + \bigg(\sigma_2^2 \mathbf{I} + \xi_2 \mathbf{H}_{22}^{H}\mathbf{Q}_{2}\mathbf{H}_{22} \bigg)^{-1} \mathbf{H}_{12}^{H}\mathbf{Q}_{1}\mathbf{H}_{12} \right|, \nonumber
\end{align}
where $ \xi_i \in (0,1) $, $ \forall i = 1,2 $ are the SI residual factors for $ \mathcal{U}_1 $ and $ \mathcal{U}_2 $, which reflect the residual SI power level after SI suppression.
On the other side, $ \mathcal{E} $ eavesdrops both confidential information from $ \mathcal{U}_1 $ and $ \mathcal{U}_2 $ simultaneously, according to two-user multiple access channel capacity results \cite{Tse_book_2005}, the sum rate of $ \mathcal{E} $ can be written as
\begin{align}
R_{e} = \log \left| \mathbf{I} + \frac{1}{\sigma_e^2} \mathbf{H}_{e,1}^{H}\mathbf{Q}_{1}\mathbf{H}_{e,1} + \frac{1}{\sigma_e^2} \mathbf{H}_{e,2}^{H}\mathbf{Q}_{2}\mathbf{H}_{e,2} \right|.
\end{align}
Thus, the achievable sum secrecy rate of this two-way transmission can be given by
\begin{align}
R_{sum} = [R_{1} + R_{2} - R_{e}]^{+},
\end{align}
where $ [*]^{+} = \max\{ *,0 \} $.
\section{Robust Sum Secrecy Rate Optimization for MIMO Two-way Full Duplex Systems}
In this section, we propose an RSSRM problem subject to the outage probability constraint of achievable sum secrecy rate and the transmit power constraints. Robust transmit solution is designed for the proposed RSSRM problem based on a \emph{moment-based} channel uncertainty model.
\subsection{Channel Uncertainty Model}\label{seciton:Moment_based_channel_uncertainty}
We assume that both $ \mathcal{U}_1 $ and $ \mathcal{U}_2 $ cannot have the perfect CSI of $ \mathcal{E} $. To account for the imperfection of $ \mathcal{E} $'s CSI, in this section, we model a moment-based random CSI error model, where the true channels at the eavesdropper can be expressed as
\begin{align}
\mathbf{H}_{e,1} = \mathbf{\bar{H}}_{e,1} + \mathbf{E}_{e,1},\label{Eev1}\\
\mathbf{H}_{e,2} = \mathbf{\bar{H}}_{e,2} + \mathbf{E}_{e,2}.\label{Eev2}
\end{align}
In \eqref{Eev1} and \eqref{Eev2}, $ \mathbf{\bar{H}}_{e,1} $ and $ \mathbf{\bar{H}}_{e,2} $ are estimated CSI of $ \mathbf{H}_{e,1} $ and $ \mathbf{H}_{e,2} $, respectively; $ \mathbf{E}_{e,1} $ and $ \mathbf{E}_{e,1} $ denote the corresponding estimated channel errors following randomly distributions, i.e., $ \textrm{vec}(\mathbf{E}_{e,i}) \sim \mathcal{D}(\mathbf{\phi}_i, \mathbf{\Omega}_i) $, $ \forall i = 1,2 $. 
In this paper, $ \mathcal{D}(\mathbf{\phi}_i, \mathbf{\Omega}_i) $ denotes an  arbitrary distribution with mean $ \mathbf{\phi}_i$ and covariance matrix $ \mathbf{\Omega}_i, \forall i $.
\begin{remark}
This channel uncertainty model adopted in this paper is more practical than the channel-error-bound model in \cite{Cuma_TVT_MIMO_Sec_2014,Zheng_Secrecy_J15,Zhengyu_Zhu_JCN_Robust_SWIPT_2016} or the completed-error-distribution-information model in \cite{Zheng_WCL_2015,Zhengyu_Zhu_Outage_Sec_TWC_2016}  since the proposed model only requires to estimate the channel error statistics.
\end{remark}
\vspace{-1em}
\subsection{Robust Sum Secrecy Rate Maximization}
Employing the channel uncertainty model described in previous section, we introduce the following optimization problem:
\small
\begin{subequations}\label{eq:Outage_SSRM}
\begin{align}
\max_{\substack{R_{s}, \mathbf{Q}_{1},\mathbf{Q}_{2}}} &~ R_{s}, \nonumber\\
&\!\!\!\!\!\!\!\!\!\!\!\!\!\!\!\!\!\!\!\!\!\!\!\! s.t.~  \min_{\substack{\textrm{vec}(\mathbf{E}_{e,1}) \sim \mathcal{D}(\mathbf{\phi}_1, \mathbf{\Omega}_1) \\ \textrm{vec}(\mathbf{E}_{e,2}) \sim \mathcal{D}(\mathbf{\phi}_2, \mathbf{\Omega}_2) }} \textrm{Pr}\{ R_{sum} \geq R_{s} \} \geq 1 - \rho,\label{eq:Outage_probability_ori}\\
&\!\!\!\!\!\!\!\!\!\! \textrm{Tr}(\mathbf{Q}_{1}) \leq P_{1},~\textrm{Tr}(\mathbf{Q}_{1}) \leq P_{2},~\mathbf{Q} \succeq \mathbf{0},~\mathbf{Q}_{2} \succeq \mathbf{0},~R_{s} \geq 0, \label{eq:Power_and_SDP_constraints}
\end{align}
\end{subequations}
\normalsize
where $\rho \in(0,1)$ is the outage probability, $P_1$ and $P_2$ are the transmit power constraints at $ \mathcal{U}_1 $ and $ \mathcal{U}_2 $, respectively.
Problem \eqref{eq:Outage_SSRM} is not convex due to the constraint \eqref{eq:Outage_probability_ori}. 
In order to make the constraint \eqref{eq:Outage_probability_ori} more tractable, we apply the first-order Taylor series approximation in $ R_{sum} $ as \cite{Cuma_TVT_MIMO_Sec_2014,Zheng_Secrecy_J15}
\begin{align}
&\!\!\!\!\! R_{sum} =  \log \left| \sigma_1^2 \mathbf{I} + \xi_1 \mathbf{H}_{11}^{H}\mathbf{Q}_{1}\mathbf{H}_{11} + \mathbf{H}_{21}^{H}\mathbf{Q}_{2}\mathbf{H}_{21} \right| \nonumber\\
& + \log \left| \sigma_1^2 \mathbf{I} + \xi_2 \mathbf{H}_{22}^{H}\mathbf{Q}_{2}\mathbf{H}_{22} + \mathbf{H}_{12}^{H}\mathbf{Q}_{1}\mathbf{H}_{12} \right| \nonumber\\
& - \log \left| \sigma_1^2 \mathbf{I} + \xi_1 \mathbf{H}_{11}^{H}\mathbf{Q}_{1}\mathbf{H}_{11} \right| - \log \left| \sigma_2^2 \mathbf{I} + \xi_2 \mathbf{H}_{22}^{H}\mathbf{Q}_{2}\mathbf{H}_{22} \right| \nonumber\\
& - \log \left| \mathbf{I} + \frac{1}{\sigma_e^2} \mathbf{H}_{e,1}^{H}\mathbf{Q}_{1}\mathbf{H}_{e,1} + \frac{1}{\sigma_e^2} \mathbf{H}_{e,2}^{H}\mathbf{Q}_{2}\mathbf{H}_{e,2} \right| \nonumber\\
& \simeq
f - \alpha + \beta - \gamma_{e,1} - \gamma_{e,2}
= \tilde{R}_{sum},
\end{align}
where
\begin{align}
f & =
\log \left| \sigma_1^2 \mathbf{I} + \xi_1 \mathbf{H}_{11}^{H}\mathbf{Q}_{1}\mathbf{H}_{11} + \mathbf{H}_{21}^{H}\mathbf{Q}_{2}\mathbf{H}_{21} \right| \nonumber\\
& + \log \left| \sigma_1^2 \mathbf{I} + \xi_2 \mathbf{H}_{22}^{H}\mathbf{Q}_{2}\mathbf{H}_{22} + \mathbf{H}_{12}^{H}\mathbf{Q}_{1}\mathbf{H}_{12} \right| \nonumber\\
& - \log \left| \sigma_1^2 \mathbf{I} + \xi_1 \mathbf{H}_{11}^{H}\mathbf{\tilde{Q}}_{1}\mathbf{H}_{11} \right| - \log \left| \sigma_2^2 \mathbf{I} + \xi_2 \mathbf{H}_{22}^{H}\mathbf{\tilde{Q}}_{2}\mathbf{H}_{22} \right| \nonumber\\
& - \frac{1}{\ln 2} \textrm{Tr} \left[ \mathbf{A}_{1} \xi_1 \mathbf{H}_{11}^{H}(\mathbf{Q}_{1} \!-\! \mathbf{\tilde{Q}}_{1})\mathbf{H}_{11} \right] \nonumber\\
& - \frac{1}{\ln 2} \textrm{Tr} \left[\mathbf{A}_{2} \xi_1 \mathbf{H}_{22}^{H}(\mathbf{Q}_{2} \!-\! \mathbf{\tilde{Q}}_{2})\mathbf{H}_{22} \right], \nonumber
\end{align}
\begin{align}
\alpha & =   \log \left| \mathbf{I} + \frac{1}{\sigma_e^2} \mathbf{\bar{H}}_{e,1}^{H}\mathbf{\tilde{Q}}_{1}\mathbf{\bar{H}}_{e,1} + \frac{1}{\sigma_e^2} \mathbf{\bar{H}}_{e,2}^{H}\mathbf{\tilde{Q}}_{2}\mathbf{\bar{H}}_{e,2} \right|, \nonumber
\end{align}
\begin{align}
\beta &= \frac{1}{\ln 2} \textrm{Tr} \left[ \mathbf{A}_{e} \bigg( \mathbf{\bar{H}}_{e,1}^{H} \mathbf{\tilde{Q}}_{1} \mathbf{\bar{H}}_{e,1} + \mathbf{\bar{H}}_{e,2}^{H} \mathbf{\tilde{Q}}_{2} \mathbf{\bar{H}}_{e,2} \bigg) \right], \nonumber\\
\gamma_{e,1} & = \frac{1}{\ln 2} \textrm{Tr} \left[ \mathbf{A}_{e} (\mathbf{\bar{H}}_{e,1} + \mathbf{E}_{e,1})^{H} \mathbf{Q}_{1} (\mathbf{\bar{H}}_{e,1}   + \mathbf{E}_{e,1}) \right], \nonumber\\
\gamma_{e,2} & = \frac{1}{\ln 2} \textrm{Tr} \left[ \mathbf{A}_{e} (\mathbf{\bar{H}}_{e,2} + \mathbf{E}_{e,2})^{H} \mathbf{Q}_{2} (\mathbf{\bar{H}}_{e,2} + \mathbf{E}_{e,2}) \right], \nonumber
\end{align}
\begin{align}
\mathbf{A}_{1} & \!=\! \bigg(\!\! \sigma_1^2 \mathbf{I} \!+\! \xi_1 \mathbf{H}_{11}^{H}\mathbf{\tilde{Q}}_{1}\mathbf{H}_{11} \!\! \bigg)^{-1}\!\!, \mathbf{A}_{2}  \!= \! \bigg(\!\! \sigma_2^2 \mathbf{I} \!+\! \xi_2 \mathbf{H}_{22}^{H}\mathbf{\tilde{Q}}_{2}\mathbf{H}_{22} \!\! \bigg)^{-1}\!\!, \nonumber
\end{align}
\begin{align}
\mathbf{A}_{e} & \!=\! \frac{1}{\sigma_e^2}\bigg( \mathbf{I} \!+\! \frac{1}{\sigma_e^2} \mathbf{\bar{H}}_{e,1}^{H}\mathbf{\tilde{Q}}_{1}\mathbf{\bar{H}}_{e,1} \!+\! \frac{1}{\sigma_e^2} \mathbf{\bar{H}}_{e,2}^{H}\mathbf{\tilde{Q}}_{2}\mathbf{\bar{H}}_{e,2} \bigg)^{-1}, \nonumber
\end{align}
also, $ \mathbf{\tilde{Q}}_{1} $ and $ \mathbf{\tilde{Q}}_{2} $ are the approximated transmit covariance matrices of $ \mathcal{U}_1 $ and $ \mathcal{U}_2 $, respectively.
In addition, since the channel errors $ \mathbf{E}_{e,1} $ and $ \mathbf{E}_{e,2} $ appear only in $ R_{e} $, the problem \eqref{eq:Outage_SSRM} can be rewritten as
	\begin{subequations}
	\begin{align}
	\max_{\mathbf{Q}_{1},\mathbf{Q}_{2}, t \geq 0} &~ f - t  \\
	& \!\!\!\!\!\!\!\!\!\!\!\!\!\!\!\! s.t. \!\!\!\! \min_{\substack{\textrm{vec}(\mathbf{E}_{e,1}) \sim \mathcal{D}(\mathbf{\phi}_1, \mathbf{\Omega}_1) \\ \textrm{vec}(\mathbf{E}_{e,2}) \sim \mathcal{D}(\mathbf{\phi}_2, \mathbf{\Omega}_2) }} \!\!\!\!\!\!\!\!\!\!\!\! \textrm{Pr} \{ \alpha \!-\! \beta \!+\! \gamma_{e,1} \!+\! \gamma_{e,2}   \!\leq\! t   \} \!\geq\! 1 \!-\! \rho, \label{eq:Outage_probability_reformulate} \\
	&\textrm{constraints}~\eqref{eq:Power_and_SDP_constraints}. \nonumber
	\end{align}
	\end{subequations}
The above problem is still non-convex due to \eqref{eq:Outage_probability_reformulate}. In order to tackle this challenge, we consider the following matrix identity:
\begin{align}\label{eq:Matrix_relations}
\textrm{Tr}(\mathbf{A}\mathbf{B}\mathbf{C}\mathbf{D}) = \textrm{vec}(\mathbf{A})^{H}(\mathbf{D}^{T} \otimes \mathbf{B}) \textrm{vec}(\mathbf{C}).
\end{align}
 By exploiting \eqref{eq:Matrix_relations}, the outage probability constraint is modified as \eqref{eq:Outage_probability_reformulation}
	\begin{align}\label{eq:Outage_probability_reformulation}
\!\!\!\!\!\!	\min_{\substack{\mathbf{e}_{e,1}  \sim  \mathcal{D}(\mathbf{\phi}_1, \mathbf{\Omega}_1) \\ \mathbf{e}_{e,2} \sim \mathcal{D}(\mathbf{\phi}_2, \mathbf{\Omega}_2) }} \!\!\!\!\!\!\!\! \textrm{Pr} \{ \mathbf{h}_{e,1}^{H} \mathbf{B}_{1} \mathbf{h}_{e,1}\!+\! \mathbf{h}_{e,2}^{H} \mathbf{B}_{2}  \mathbf{h}_{e,2}  \!\leq\! (t \!+\! \beta \!-\! \alpha) \ln 2  \} \!\geq \! 1 \!-\! \rho,
	\end{align}
where
\begin{align}
 \mathbf{B}_{i} &= \mathbf{A}_{e}^{T} \otimes \mathbf{Q}_{i},~
 \mathbf{h}_{e,i} = \mathbf{\bar{h}}_{e,i} + \mathbf{e}_{e,i},\nonumber\\
 \mathbf{\bar{h}}_{e,i} & = \textrm{vec}(\mathbf{\bar{H}}_{e,i}),~ \mathbf{e}_{e,i} = \textrm{vec}(\mathbf{\bar{E}}_{e,i}), ~\forall i = 1,2. \nonumber
\end{align}
The outage probability constraint in \eqref{eq:Outage_probability_reformulation} neither is convex nor has simple closed form. To tackle the problem, in the following, we replace the constraint by two convex approximations  based on a \emph{Markov's inequality} \cite{Xin_He_Random_CSI_MIMO_TWC_2014} and Semidefinite programming (SDP).
\subsection{Markov-Inequality Approach}\label{section Markov_inequality} Based on a Markov's inequality \cite{Xin_He_Random_CSI_MIMO_TWC_2014}, we introduce a simple lower bound for the minimum outage probability, i.e., the left hand side of \eqref{eq:Outage_probability_reformulation}, as follows:
\begin{align}\label{eq:Markove_inequality_lower_bound}
&\!\!\!\! \min_{\substack{\mathbf{e}_{e,1} \sim \mathcal{D}(\mathbf{\phi}_1, \mathbf{\Omega}_1) \\ \mathbf{e}_{e,2} \sim \mathcal{D}(\mathbf{\phi}_2, \mathbf{\Omega}_2) }} \textrm{Pr} \{ \mathbf{h}_{e,1}^{H} \mathbf{B}_{1} \mathbf{h}_{e,1} + \mathbf{h}_{e,2}^{H} \mathbf{B}_{2} \mathbf{h}_{e,2} \leq ( t + \beta - \alpha ) \ln 2 \} \nonumber\\
&  \geq  1 - \frac{\mathbb{E}\{ \mathbf{h}_{e,1}^{H} \mathbf{B}_{1} \mathbf{h}_{e,1} +  \mathbf{h}_{e,2}^{H} \mathbf{B}_{2} \mathbf{h}_{e,2} \}}{ ( t + \beta - \alpha ) \ln 2 } \nonumber\\
& = 1 - \frac{\textrm{Tr}(\mathbf{B}_{1}\mathbf{\Gamma}_{1} + \mathbf{B}_{2}\mathbf{\Gamma}_{2})}{ ( t + \beta - \alpha ) \ln 2 },
\end{align}
where $ \mathbf{\Gamma}_{1} = \mathbf{\Omega}_{1} + (\mathbf{\bar{h}}_{e,1} + \mathbf{\phi}_{1})(\mathbf{\bar{h}}_{e,1} + \mathbf{\phi}_{1})^{H} $, and $ \mathbf{\Gamma}_{2} = \mathbf{\Omega}_{2} + (\mathbf{\bar{h}}_{e,2} + \mathbf{\phi}_{2})(\mathbf{\bar{h}}_{e,2} + \mathbf{\phi}_{2})^{H} $.
Substituting the lower bound in \eqref{eq:Markove_inequality_lower_bound} into the left hand side of \eqref{eq:Outage_probability_reformulation} with some mathematical manipulations, the problem in \eqref{eq:Outage_SSRM} can be reformulated as:
\begin{align}\label{eq:Outage_SSM_results_markov_inequality}
\max_{\mathbf{Q}_{1},\mathbf{Q}_{2},t \geq 0} &~ f - \alpha + \beta - t \nonumber\\
s.t. &~ \textrm{Tr}(\mathbf{B}_{1}\mathbf{\Gamma}_{1} + \mathbf{B}_{2}\mathbf{\Gamma}_{2}) \leq  ( t + \beta - \alpha )  \rho \ln 2, \nonumber\\
&\textrm{constraints}~\eqref{eq:Power_and_SDP_constraints}.
\end{align}
For given approximated transmit covariance matrices $ \mathbf{\tilde{Q}}_{1} $ and $ \mathbf{\tilde{Q}}_{2} $, problem \eqref{eq:Outage_SSM_results_markov_inequality} is convex with respect to $ \mathbf{{Q}}_{1} $ and $ \mathbf{{Q}}_{2} $. Hence, it can be effectively solved by interior-point methods \cite{boyd_B04}. The question raised here is how to obtain the values for $ \mathbf{\tilde{Q}}_{1} $ and $ \mathbf{\tilde{Q}}_{2} $. To that end, we adopt DC programming to obtain optimal transmit covariance matrices $ \mathbf{{Q}}_{1} $ and $ \mathbf{{Q}}_{2} $ as follows. We first randomly generate $ \mathbf{\tilde{Q}}_{1} $ and $ \mathbf{\tilde{Q}}_{2} $ and use those matrices to solve \eqref{eq:Outage_SSM_results_markov_inequality} to attain $ \mathbf{{Q}}_{1} $ and $ \mathbf{{Q}}_{2} $. The newly attained $ \mathbf{{Q}}_{1} $ and $ \mathbf{{Q}}_{2} $ will be assign as $ \mathbf{\tilde{Q}}_{1} $ and $ \mathbf{\tilde{Q}}_{2} $  to be used at the next iteration. The process is repeated until a stationary solution is achieved.

\subsection{Semidefinite-Programming (SDP) Approach}
It is worth mentioning that the outage probability in \eqref{eq:Outage_probability_reformulation} is a worst-case probability with quadratic inequality, which can be equivalently reformulated as a convex conic framework. Here, we propose a SDP approach to provide a convex approximation of the minimum outage probability \eqref{eq:Outage_probability_reformulation}. We start developing our SDP approach by introducing the following theorem.
\begin{theorem}\label{lemma Distributionall_robust_lemma}
 The constraint \eqref{eq:Outage_probability_reformulation} can be safely approximated as:
 \begin{subequations}\label{eq:Distributionally_robust_SDP_solution}
 \begin{align}
 \min_{\substack{\mu \in \mathbb{R}\\ \mathbf{M} \in \mathbb{H}^{N}}} &~ \mu + \rho^{-1} \textrm{Tr}(\mathbf{\Pi}\mathbf{M}) \leq 0 \label{eq:Distributionally_1}\\
 s.t. &~ \mathbf{M} \succeq \left[ \begin{array}{cc}
 \mathbf{B} & \mathbf{0}_{1} \\
\mathbf{0}_{1}^{H} & - ( t + \beta - \alpha ) \ln 2  - \mu
 \end{array}\right], \label{eq:Distributionally_2}\\
 &~ \mathbf{M} \succeq \mathbf{0}, \label{eq:Distributionally_3}
 \end{align}
 \end{subequations}
 where
 \begin{align}
 & \mathbf{B} = \left[\begin{array}{cc}
 \mathbf{B}_{1}  & \mathbf{0}_{2}  \\
 \mathbf{0}_{2}^{H} & \mathbf{B}_{2}
 \end{array}\right],\nonumber\\
 & \mathbf{\Pi} = \left[\begin{array}{ccc}
 \mathbf{\Omega}_{1} &  \mathbf{0}_{2}  &  \mathbf{0}_{3} \\
 \mathbf{0}_{2}^{H} & \mathbf{\Omega}_{2} & \mathbf{0}_{4} \\
 \mathbf{0}_{3}^{H} & \mathbf{0}_{4}^{H}  & 0
 \end{array}\right] + \left[\begin{array}{ccc}
 \mathbf{\bar{h}}_{e,1} + \mathbf{\phi}_{1} \\
 \mathbf{\bar{h}}_{e,2} + \mathbf{\phi}_{2} \\
 1
 \end{array}\right] \left[\begin{array}{ccc}
 \mathbf{\bar{h}}_{e,1} + \mathbf{\phi}_{1} \\
 \mathbf{\bar{h}}_{e,2} + \mathbf{\phi}_{2} \\
 1
 \end{array}\right]^{H}, \nonumber
 \end{align}
 $ N = N_{E}(N_{T,1}+N_{T,2})+1 $, $ \mathbf{0}_{1} =  \mathbf{0}_{N_{E}(N_{T,1}+N_{T,2}) \times 1} $, $ \mathbf{0}_{2} = \mathbf{0}_{N_{T,1}N_{E} \times N_{T,2}N_{E}} $, $ \mathbf{0}_{3} = \mathbf{0}_{N_{T,1}N_{E} \times 1} $, and $ \mathbf{0}_{4} = \mathbf{0}_{N_{T,2}N_{E} \times 1} $.
\end{theorem}
\begin{IEEEproof}
Please refer to Appendix.
\end{IEEEproof}
Exploiting \emph{Theorem} \ref{lemma Distributionall_robust_lemma}, the proposed RSSRM problem can be written as
\begin{subequations}\label{eq:Distributionally_SDP_results}
\begin{align}
\max_{\mathbf{Q}_{1},\mathbf{Q}_{2}, R_{s}, \mu, \mathbf{M}, t \geq 0} &~ f  - t \nonumber\\
s.t. &~  \mu + \rho^{-1} \textrm{Tr}(\mathbf{\Pi}\mathbf{M}) \leq 0, \label{eq:Distributionally_1_results}\\
&\textrm{constraints}~ \eqref{eq:Power_and_SDP_constraints},~\eqref{eq:Distributionally_2},~\eqref{eq:Distributionally_3}.
\end{align}
\end{subequations}
It is easily observed that \eqref{eq:Distributionally_1_results} holds if and only if there exists a feasible point $(\mu \in \mathbb{R}, \mathbf{M} \in \mathbb{H}^{N}) $ in the minimum outage probability constraint \eqref{eq:Distributionally_1} such that $ \mu + \rho^{-1} \textrm{Tr}(\mathbf{\Pi}\mathbf{M}) \leq 0 $. For given the approximated transmit covariance matrices $ \mathbf{\tilde{Q}}_{1} $ and $ \mathbf{\tilde{Q}}_{2} $, problem \eqref{eq:Distributionally_SDP_results} is convex with respect to $ \mathbf{{Q}}_{1} $ and $ \mathbf{{Q}}_{2} $. 
Therefore, a similar procedure adopting the DC programming, as described in the previous section, is performed to iteratively update the stationary solution to the RSSRM problem \eqref{eq:Outage_SSRM}.

\subsection{Performance Analysis}
In terms of complexity, by comparing Markov-inequality approach, i.e., problem \eqref{eq:Outage_SSM_results_markov_inequality}, and the SDP approach, i.e., problem \eqref{eq:Distributionally_SDP_results}, it is easily observed that the former has a lower computation complexity than the latter.

In terms of tightness, the following lemma reveals the relative tightness of the two proposed approaches.
\begin{lemma}\label{tightness} Every feasible solution to  \eqref{eq:Outage_SSM_results_markov_inequality} is also a feasible solution to \eqref{eq:Distributionally_SDP_results}.
\end{lemma}
\begin{IEEEproof}
Assuming $ (\mathbf{\hat{Q}}_{1},\mathbf{\hat{Q}}_{2}, \hat{t}) $ is a feasible solution to the problem \eqref{eq:Outage_SSM_results_markov_inequality}. Substitute $ (\mathbf{\hat{Q}}_{1},\mathbf{\hat{Q}}_{2}, \hat{t}) $ into \eqref{eq:Distributionally_SDP_results}, we have the following solution,
\begin{align}\label{eq:Feasible_solution}
\mathbf{\hat{M}}= \left[\begin{array}{ccc}
\mathbf{\hat{B}}_{1} & \mathbf{0}_{2} & \mathbf{0}_{3} \\
\mathbf{0}_{2}^{H} & \mathbf{\hat{B}}_{2} & \mathbf{0}_{4} \\
\mathbf{0}_{3}^{H} & \mathbf{0}_{4}^{H} & 0
\end{array}\right],~ \hat{\mu} = - (\hat{t} + \beta - \alpha) \ln2,
\end{align}
where $ \mathbf{\hat{B}}_{1} = \mathbf{A}_{e}^{T} \otimes \mathbf{\hat{Q}}_{1} $ and $ \mathbf{\hat{B}}_{2} = \mathbf{A}_{e}^{T} \otimes \mathbf{\hat{Q}}_{2} $. From \eqref{eq:Feasible_solution}, one can verify that $(\mathbf{\hat{Q}}_{1},\mathbf{\hat{Q}}_{2},\mathbf{\hat{M}},\hat{t},\hat{\mu}) $ is also a feasible solution to the problem  \eqref{eq:Distributionally_SDP_results}. It is worth mentioning that a feasible solution to \eqref{eq:Distributionally_SDP_results} may be infeasible to  \eqref{eq:Outage_SSM_results_markov_inequality}.
\end{IEEEproof}
\begin{remark}\label{remark2}
Lemma \ref{tightness} indicates that the SDP approach provides a tighter approximation of the proposed RSSRM problem \eqref{eq:Outage_probability_reformulation} than the Markov-inequality approach does.
\end{remark}
\section{Simulation Results}
In this section, simulation results are provided to validate the performance of our proposed robust RSSRM approaches, i.e., the proposed robust Markov-inequality and SDP approaches. We also compare the performance of the proposed approaches against that of the FD scheme based on perfect eavesdropper's CSI, and half-duplex (HD) schemes adopting Markov-inequality and SDP approaches.

We consider the secure MIMO two-way FD system that consists of two multi-antenna legitimate nodes, i.e., $ \mathcal{U}_{1} $ and $ \mathcal{U}_{2} $, and one multi-antenna eavesdropper, i.e., $ \mathcal{E} $. It is assumed that both $ \mathcal{U}_{1} $ and $ \mathcal{U}_{2} $ are equipped with five transmit antennas, i.e., $ N_{T,1} = N_{T,2} = 5 $, and two receive antennas, i.e., $ N_{R,1} = N_{R,2} = 2 $, whereas $ \mathcal{E} $ consists of two antennas, i.e., $ N_{E} = 2 $. Also, the noise variance matrices at three nodes $ \mathcal{U}_{1} $, $ \mathcal{U}_{2} $, and $ \mathcal{E} $ are set to be $ \mathbf{I}$, i.e., $ \sigma_{i}^{2} =  1,~\forall i = 1,2,e $. Without any loss of generality, we assume that both $ \mathcal{U}_{1} $ and $ \mathcal{U}_{2} $ have the same FD SI residual factor $ \xi_1 = \xi_2 = 0.01 $, and the same maximum available transmit power $ P_{1} = P_{2} =P= 5~\textrm{dB} $. All channel coefficients are generated as circularly symmetric independent and identically distributed Gaussian random variables with zero-mean and unit variance. In addition, the channel error mean are set to be zero-mean (i.e., $ \mathbf{\phi}_{i} = 0 $, $ \forall i = 1,2 $), and the channel error covariance matrix to be $ \Omega_{i} = \varepsilon_{i} \mathbf{I}$, where $ \varepsilon_{i} = \varepsilon = 0.005,~\forall i = 1,2 $ unless otherwise stated. The outage probability threshold is set to be $ \rho = 0.05 $.\\
\begin{figure}[!htbp]
	\centering
	\includegraphics[scale = 0.47]{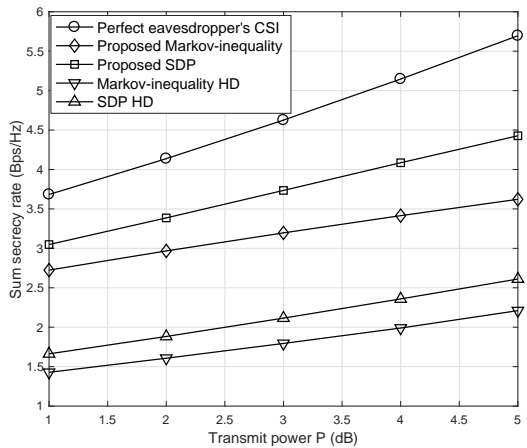}
	\caption{Sum secrecy rate vs transmit power $P$.}
	\label{fig:OSSR_vs_Power}
\end{figure}
\indent In Fig.~\ref{fig:OSSR_vs_Power}, we evaluate the sum secrecy rate performances of several approaches versus the maximum available transmit power  $ P $. From the figure, it is clear that increasing maximum available transmit power  leads to an increase of the sum secrecy rate. It can be observed that the SDP approach outperforms the Markov-inequality approach in terms of providing higher sum secrecy rate. This is due to the fact that the former employs a tighter convex approximation than the latter does. In other words, the Markov-inequality approach is more conservative than the SDP counterpart. This confirms Lemma \ref{tightness} and the statement in Remark \ref{remark2}.

Fig.~\ref{fig:OSSR_vs_Power} shows that having perfect eavesdropper's CSI results in the highest sum secrecy rate. However, due to the nature of eavesdroppers, their CSI are normally outdated or even hardly to obtain in practice. Hence, perfect eavesdropper's CSI is an impractical assumption.  
\begin{figure}[!htbp]
	\centering
	\includegraphics[scale = 0.47]{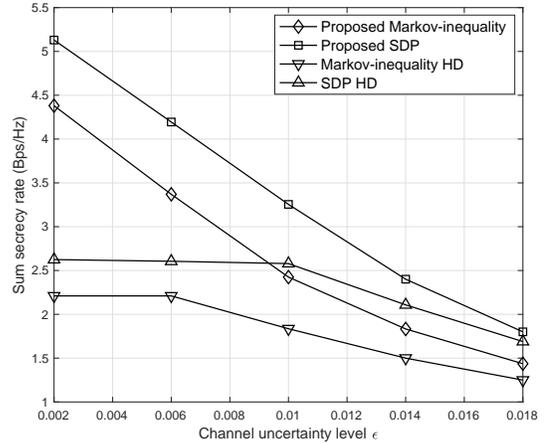}
	\caption{Sum secrecy rate vs channel uncertainty level $\varepsilon$.}
	\label{fig:OSSR_vs_Error}
\end{figure}

In Fig.~\ref{fig:OSSR_vs_Error}, the sum secrecy rate is evaluated with different channel uncertainty levels $ \varepsilon $. It can be seen from the figure that the sum secrecy rate decreases when the accuracy of CSI estimation decreases, i.e. $ \varepsilon $ increases. Moreover, the performance of the SDP approach prevails that of the Markov-inequality approach. Here, the statement in Remark \ref{remark2} is verified again. The results shown in Figs.~\ref{fig:OSSR_vs_Power} and \ref{fig:OSSR_vs_Error} indicate that FD approaches provide significant improvements in the sum secrecy rate compared with their HD counterparts.

\begin{figure}[!htbp]
	\centering
	\includegraphics[scale = 0.47]{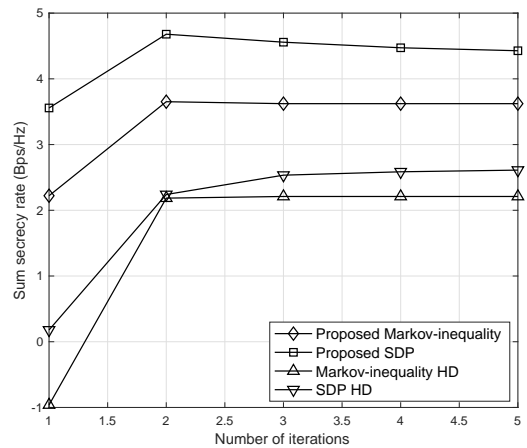}
	\caption{Sum secrecy rate vs  number of iterations.}
	\label{fig:OSSR_vs_Iteration}
\end{figure}

Finally, we evaluate the convergence performance of our proposed schemes in Fig. \ref{fig:OSSR_vs_Iteration}. It is observed from this figure that our proposed schemes converge to a stationary solution in terms of the sum secrecy rate in just 3 iterations.
\section{Conclusions}
This paper investigated MIMO FD two-way secrecy systems. Taking into the consideration a random distributed eavesdropper's CSI model, we formulated an RSSRM problem subject to the outage probability constraint of the achievable sum secrecy rate and the transmit power constraint. We proposed two approaches based on a Markov-inequality and SDP to tackle this RSSRM problem. Moreover, DC approximation is employed to iteratively optimize the transmit covariance matrices of both legitimate nodes. Simulation results showed that the proposed FD approaches outperform the associated HD schemes, also, the SDP approach has a better performance than the Markov-inequality approach in terms of higher achievable sum secrecy rate.
\vspace{-0.5em}
\section*{Appendix}
\vspace{-0.5em}
In order to prove \emph{Theorem} \ref{lemma Distributionall_robust_lemma}, we consider the following \emph{lemmas}:
\begin{lemma}
	\cite[Theorem 2.2]{Zymler_Distributionally_2013} Assuming that $ f(x) $ is a quadratic function (i.e., concave) with respect to $ x $, where $ f(x): \mathbb{C}^{\bar{N}} \rightarrow \mathbb{R} $ is a continuous function. Then the following relation hold:
	\begin{align}
	\sup_{x \sim \mathcal{D}(\mathbf{\phi},\mathbf{\Omega})} \!\!\!\!\! \textrm{CVaR}_{\rho} (f(x)) \leq 0 \Leftrightarrow \!\!\!\!\! \inf_{x \sim \mathcal{D}(\mathbf{\phi},\mathbf{\Omega})} \!\!\!\! \textrm{Pr}(f(x) \leq 0) \geq 1 - \rho,
	\end{align}
	where $ 0 \leq \rho \leq 1 $, and $ \textrm{CVaR}_{\rho} $ is the Conditional Value-at-Risk function that can be expressed as
	\begin{align}\label{eq:Definition_CVaR}
	\textrm{CVaR}_{\rho}(f(x)) = \inf_{\mu \in \mathbb{R}} \bigg[ \mu + \frac{1}{\rho} \mathbb{E}_{x} [ f(x) - \mu ]^{+} \bigg].
	\end{align}
\end{lemma}
	\begin{lemma}\label{lemma Distributionally_lemma2}
   \cite[Lemma A.1]{Zymler_Distributionally_2013} If $ f:\mathbb{C}^{\bar{N}} \rightarrow \mathbb{R} $ is continuous function, the worst-case expectation can be defined as follows:
 	\begin{align}
	\sup_{x \sim \mathcal{D}(\mathbf{\phi},\mathbf{\Omega})} \mathbb{E}_{x} [(f(x))^+] &= \inf_{M \in \mathbb{H}^{N+1}, \mathbf{M} \succeq \mathbf{0}} \textrm{Tr}(\mathbf{\Gamma} \mathbf{M}) \nonumber\\
	s.t. &~\left[\!\!\begin{array}{cc}
	x^{H} \!\!&\!\!  1
	\end{array}\!\!\right] \mathbf{M} \left[\!\!\begin{array}{cc}
	x \\ 1
	\end{array}\!\!\right] \!\geq \! f(x),~\forall x \in \mathbb{C}^{\bar{N}},
	\end{align}
	where
	\begin{align}
  \mathbf{\Gamma} = \left[\begin{array}{cc}
	\mathbf{\varSigma} + \mathbf{\varpi} \mathbf{\varpi}^{H} & \mathbf{\varpi} \\
	\mathbf{\varpi}^{H} & 1
	\end{array}\right].
	\end{align}
	\end{lemma}
	Now, we apply both above \emph{lemmas} in the minimum outage probability constraint \eqref{eq:Outage_probability_reformulation}. Let $ g(\mathbf{h}_{e,i},\mathbf{Q}_{i}) = \mathbf{h}_{e,1}^{H} \mathbf{B}_{1} \mathbf{h}_{e,1} + \mathbf{h}_{e,2}^{H} \mathbf{B}_{2}  \mathbf{h}_{e,2} - t \ln 2 , \forall i = 1,2 $, we have
	\begin{align}\label{eq:Distributionally_CVaR_derivations}
	\eqref{eq:Outage_probability_reformulation} \Rightarrow \sup_{\mathbf{e}_{e,i} \sim \mathcal{D}(\mathbf{\phi}_{i}, \mathbf{\Omega}_{i})} \textrm{CVaR}_{\rho} (g(\mathbf{h}_{e,i},\mathbf{Q}_{i})) \leq 0,~\forall i = 1, 2.
	\end{align}
According to the definition of $ \textrm{CVaR}_{\rho} $ in \eqref{eq:Definition_CVaR},
\begin{align}\label{eq:CVaR_derive}
& \sup_{\mathbf{e}_{e,i} \sim \mathcal{D}(\mathbf{\phi}_{i}, \mathbf{\Omega}_{i})} \textrm{CVaR}_{\rho} (g(\mathbf{h}_{e,i},\mathbf{Q}_{i})) \nonumber\\
& = \sup_{\mathbf{e}_{e,i} \sim \mathcal{D}(\mathbf{\phi}_{i}, \mathbf{\Omega}_{i})} \inf_{\mu \in \mathbb{R}} \bigg\{ \mu + \frac{1}{\rho} \mathbb{E} [g(\mathbf{h}_{e,i},\mathbf{Q}_{i}) - \mu]^{+} \bigg\} \nonumber\\
& = \inf_{\mu \in \mathbb{R}} \bigg\{ \mu + \frac{1}{\rho} \sup_{\mathbf{e}_{e,i} \sim \mathcal{D}(\mathbf{\phi}_{i}, \mathbf{\Omega}_{i})} \mathbb{E} [g(\mathbf{h}_{e,i},\mathbf{Q}_{i}) - \mu]^{+} \bigg\}.
\end{align}
From the derivations of \eqref{eq:CVaR_derive}, the maximization and minimization operations have been interchanged, which has been justified by a stochastic saddle point theorem \cite{Shapiro_OMS_2002,Zymler_Distributionally_2013}. By exploiting \emph{Lemma} \ref{lemma Distributionally_lemma2}, the supremum in \eqref{eq:CVaR_derive} can be equivalently modified as
\begin{subequations}
\begin{align}
\inf_{\mathbf{M} \in \mathbb{H}^{N}, \mathbf{M} \succeq \mathbf{0}} & ~ \textrm{Tr}(\mathbf{\Pi} \mathbf{M}) \\
&\!\!\!\!\!\!\!\!\!\!\!\!\!\!\!\!\! s.t.~  \left[\!\!\begin{array}{ccc}
\mathbf{h}_{e,1}^{H} \!\!&\!\! \mathbf{h}_{e,2}^{H} \!\!&\!\! 1
\end{array}\!\!\right] \mathbf{M} \left[\!\!\begin{array}{ccc}
\mathbf{h}_{e,1} \\ \mathbf{h}_{e,2} \\ 1
\end{array}\!\!\right] \geq  \mathbb{E} [g(\mathbf{h}_{e,i},\mathbf{Q}_{i}) - \mu],\nonumber\\ &~\forall i = 1,2, \label{eq:Quadratic_function_reformulation}
\end{align}
\end{subequations}
where $ \mathbf{\Pi} $ and $ N $ have been defined in \eqref{eq:Distributionally_robust_SDP_solution}. Also, it is easily verified that $ g(\mathbf{h}_{e,i},\mathbf{Q}_{i}) $ is quadratic function with respect to $ \mathbf{h}_{e,i} $, thus, the constraint \eqref{eq:Quadratic_function_reformulation} holds if and only if
\begin{align}\label{eq:Distributionally_M}
\mathbf{M} \succeq  \left[\begin{array}{ccc}
\mathbf{B}_{1} & \mathbf{0}_{2} & \mathbf{0}_{3} \\
\mathbf{0}_{2}^{H} & \mathbf{B}_{2} & \mathbf{0}_{4} \\
\mathbf{0}_{3}^{H} & \mathbf{0}_{4}^{H} & - ( t + \beta - \alpha ) \ln 2 - \mu
\end{array}\right].
\end{align}
Hence, according to \eqref{eq:Distributionally_CVaR_derivations}, \eqref{eq:CVaR_derive} and \eqref{eq:Distributionally_M}, we complete the proof of \emph{Theorem} \ref{lemma Distributionall_robust_lemma}.
\bibliographystyle{ieeetr}
\bibliography{my_references}

\end{document}